\documentclass[prb,twocolumn,showpacs,amsmath]{revtex4}

\usepackage{graphicx}
\usepackage[dvips]{color}
 
\begin{document}
\title{Phonon Density of States in Nd(O$_{1-x}$F$_{x}$)FeAs}

\author{M. Le Tacon, M. Krisch, A. Bosak}
\affiliation{European Synchrotron Radiation Facility, BP 220, F-38043 Grenoble Cedex, France}

\author{J.-W. G. Bos, S. Margadonna}
\affiliation{School of Chemistry, University of Edinburgh, West Mains Road, EH9 3JJ Edinburgh, UK.}
\date{\today}

\begin{abstract}
We report measurements of the phonon density-of-states in iron oxypnictide superconductors by inelastic x-ray scattering. A good agreement with ab-initio calculations that do not take into account strong electronic correlations is found, and an unpredicted softening of phonon branches under F doping of these compounds is observed. Raman scattering experiments lead us to conclude that this softening is not related to zone center phonons, and consequently imply an important softening of the relevant phonon branches at finite momentum transfer Q.
\end{abstract}

\pacs{74.25.Kc, 78.70.Ck, 63.20.-e}

\maketitle
\date{\today}

\par
In the last few months, the discovery of a new family of iron-based oxypnictides superconductors~\cite{Kamihara_JACS2008, Takahashi_Nature2008} has been of great interest to the community, in particular because of their high critical temperatures, that now reach up to 55 K in Sm$_{0.9}$F$_{0.1}$FeAs~\cite{Ren_CPL2008}. 
Their study is also motivated by the striking similarities of these new compounds with high-$T_c$ cuprates (layered structure, doping-induced superconductivity, strong electronic correlations, proximity of magnetic phases ...), for which there is still no consensus on the origin of superconductivity despite 20 years of intense research.
The origin of superconductivity in oxypnictides is already controversial. Experimentally, either conventional superconducting properties (e.g. a BCS-like gap\cite{Chen_Nature2008}) or, on the contrary, occurrence of strong coupling values for the superconducting parameters~\cite{Liu_condmat2008} are found. 
~On the other hand, on theoretical grounds, serious arguments based on the strongly correlated character of these materials suggest a more unconventional superconductivity mechanism~\cite{Haule_PRL2008}. According to recent \textit{ab-initio} lattice dynamical calculations~\cite{PRL_Singh2008, Boeri}, the Migdal-Eliashberg parameters $\lambda$ and $\omega_n$ cannot explain superconductivity above roughly 1 K in this class of compounds. A pure electron-phonon coupling as a mechanism for superconductivity is thus likely excluded, but still, as in superconducting cuprates where a variety of unusual behaviours of the lattice vibrations have been reported, studying lattice dynamics in oxypnictides can provide valuable information. To date, only polycrystalline oxypnictides samples are available, therefore preventing any complete phonon dispersion studies. In this paper, we determine experimentally the phonon density-of-states (PDOS) of the compound Nd(O$_{1-x}$F$_{x}$)FeAs (x = 0.0 and 0.15) using inelastic x-ray scattering (IXS).

\par
\par
\begin{figure}[ptbh]
\begin{center}
\includegraphics[width=0.92\columnwidth]{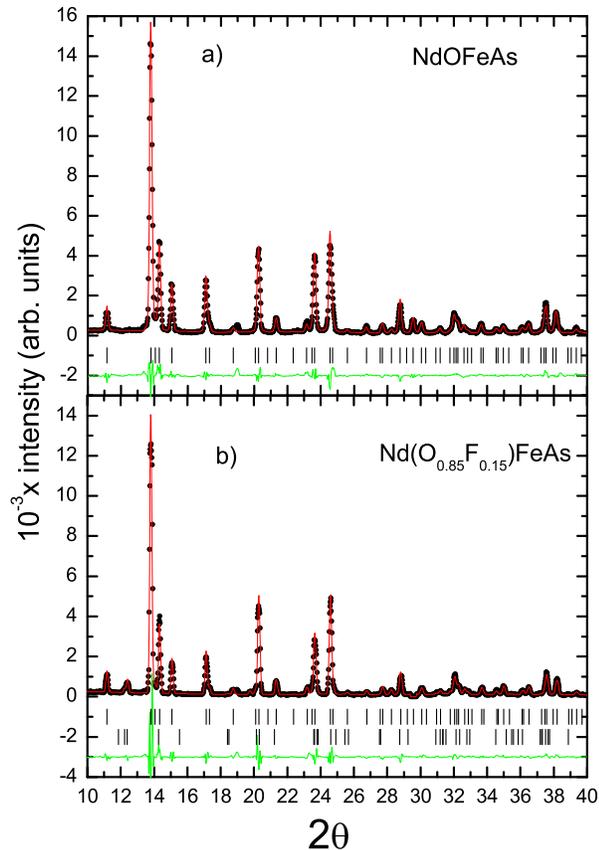}
\end{center}\vspace{-5mm}
\caption{(Color online) a) measured (dots) and calculated (solid line) synchrotron X-ray powder diffraction profiles ($\lambda$=0.7 $\AA$) for NdOFeAs at room temperature. The lower solid lines show the difference profiles and the tick marks show the reflection positions. b) measured (dots) and calculated (solid line) synchrotron X-ray powder diffraction profiles ($\lambda$=0.7 $\AA$) for NdO$\-(0.85)$F$\-(0.15)$FeAs at room temperature. The lower solid lines show the difference profiles and the tick marks show the reflection positions (from top to bottom NdO$\-(0.85)$F$\-(0.15)$FeAs, NdOF).
}
\label{fig1}
\end{figure}

Polycrystalline NdOFeAs and NdO$_{0.85}$F$_{0.15}$FeAs samples were prepared by a two-step solid state
reaction method. Stoichiometric amounts of NdAs (prepared by reacting Nd and As at 900 \ensuremath{^\circ}C for
12 hours), Fe, Fe$_2$O$_3$ and NdF$_3$ were homogenized using mortar and pestle and pressed into pellets
using an 8 mm steel dye and a 10 ton press. The pellets were inserted in tantalum cells and then
heated in vacuum sealed quartz tubes for 2 $\times$ 24 hours at 1150 \ensuremath{^\circ}C with an intermediate
homogenization.
X-ray powder diffraction measurements were performed at $\lambda$ = 0.7$\AA$ on the Swiss Norwegian Beamline at the European Synchrotron Radiation Facility (ESRF).
The GSAS suite of programs and EXPGUI graphical user interface were used for Rietveld fitting~\cite{GSAS1, GSAS2}. The results are shown in Figure~\ref{fig1}. No impurity phases were found in the parent compound while 5\% of the NdOF phase were found in the doped sample.

\par
The IXS experiment was carried out on beamline ID28 at the ESRF using the silicon (999) configuration at 17.794 keV with an instrumentation energy resolution of 3.0 meV. 
Within the limit of single phonon scattering (Stokes process), the intensity of the inelastically scattered signal with energy (momentum) transfer $\omega$ ($\textbf{Q}$) is directly proportional to the dynamical structure factor~\cite{Burkel_RPP99}:

\begin{eqnarray}
\label{SofQ}
\nonumber
S(\textbf{Q},\omega) & =& \sum_j \bigg | \sum_n \frac{f_n(\textbf{Q})e^{i\textbf{Q}.\textbf{r$_n$} - W_n}(\textbf{Q}.\hat \sigma_n(\textbf{Q}, j))}{\sqrt{M_n}}\bigg |^2\\
&  & \times \frac{(1+n(\omega_{\textbf{Q}, j},T))\delta(\omega-\omega_{\textbf{Q}, j})}{\omega_{\textbf{Q}, j}}\\
\nonumber
\end{eqnarray}

where $f_n(\textbf{Q})$ is the atomic form factor of the $n^{th}$ atom (at position $\bf{r_n}$) in the unit cell, $M_n$ its mass and $W_n$ the Debye-Waller factor. The summation index $j$ refers to the $j^{th}$ phonon branch and $\hat \sigma_n(\textbf{Q}, j)$ and $\omega_{\textbf{Q}, j}$ to the associated eigenvector and eigenvalues,  respectively.
Finally, $n(\omega, T)=(exp(\omega/k_B T)-1)^{-1}$ stands for the Bose factor. The results presented here have all been obtained at room temperature.
In the case of a powder, the dynamical structure factor must be averaged over the sphere of radius $Q = \|\textbf{Q}\|$, and, as previously demonstrated~\cite{Bosak_PRB2005}, averaging several measurements over a large range of $Q$ ($Q_{min} < Q < Q_{max})$ gives access to a quantity close to the generalized PDOS.
On ID28, the scattered photons are analysed by a set of 8 analysers that have been centered on two different set of angles to cover a $Q$ range from 53 nm$^{-1}$ to 73 nm$^{-1}$ ($\langle Q \rangle =$ 63 $nm^{-1}$).
~In this case, the dynamical structure factor reduces to

\begin{eqnarray}
\label{SofQ2}
\nonumber
S(Q_{min}, Q_{max},\omega) & =& A\langle Q \rangle^2 (1+n(\omega,T))\\
 & & \times \sum_n\frac{e^{-2W_n} f_n(\langle Q \rangle)^2 G_n(\omega)}{M_n \omega}
\end{eqnarray}

In reality, the single phonon scattering limit does not hold, and multiphonon contributions to the IXS response has to be taken into account. Although it is clear that the recursive procedure described in refs.~\onlinecite{Kohn_HI2000, Bosak_PRB2005} is only exact in the case of monoatomic systems, it enables us to subtract most of this multiphonon contribution from our data.

\begin{figure}[ptbh]
\begin{center}
\includegraphics[width=0.9\columnwidth]{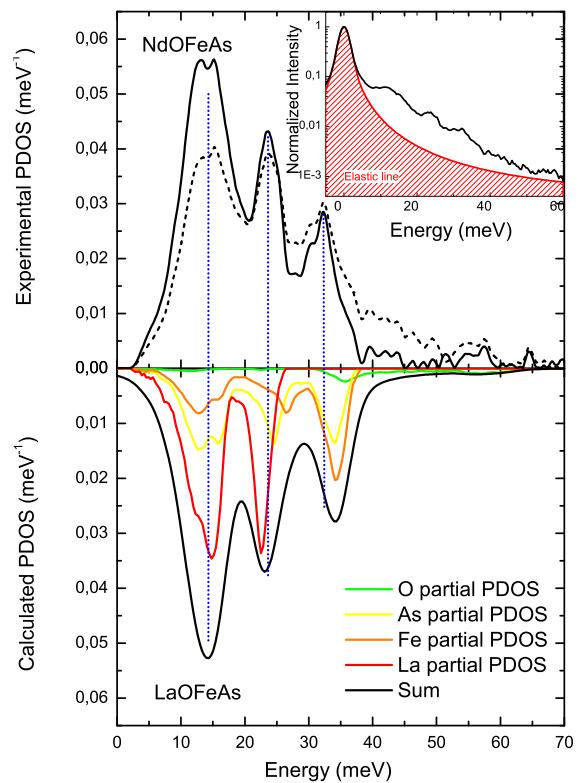}
\end{center}\vspace{-5mm}
\caption{(Color online) Comparison of the generalized phonon density-of-states of NdOFeAs (upper panel) with the results of a density functional pertubation theory calculation (lower panel - see details in the text). The dashed curve in the upper panel corresponds to the quantity $I_{inelastic}(\omega)\times [\omega/(1+n(\omega,T))]$, before multiphonon contribution subtraction.
The areas have been normalized to one.
The inset of the upper panel shows the IXS measurement prior to the subtraction of the elastic line.}
\label{fig2}
\end{figure}

In the upper panel of Figure~\ref{fig2}, we show the quantity $I_{inelastic}(\omega)\times [\omega/(1+n(\omega,T))]$ (dashed lines), where the inelastic signal $I_{inelastic}(\omega)$ is obtained after summation of individual spectra weighted by the analyser efficiencies and subtraction of the elastic contribution (see inset of Figure~\ref{fig2}), as well as the generalized PDOS obtained after multiphonon subtraction for pure NdOFeAs. 
The broad feature extending up to 65 meV in $I_{inelastic}(\omega)\times [\omega/(1+n(\omega,T))]$ is essentially removed when removing the multiphonon contribution, and in the remaining spectra, we observe essentially 3 peaks of about 5-6 meV full-width-half-maximum located around 13, 24 and 32 meV.
In the lower panel of Figure~\ref{fig2}, we report the theoretical results obtained by Boeri et al.~\cite{Boeri} on a similar compound, LaOFeAs. 
To allow a direct comparison between the experimental data and the theoretical calculation, we need to weight each calculated partial density-of-states $G_n(\omega)$ by its corresponding factor $e^{-2W_n} f_n(\langle Q \rangle)^2 M_n^{-1}$ (see table~\ref{tab:param}).

The Debye-Waller factors we used have been determined experimentally by high resolution x-ray diffraction on samples from the same batch (see details in ref. \onlinecite{Bos_condmat} - note that in first approximation, we have chosen to apply to La in the calculation the factor experimentally obtained for Nd). The result convoluted by the experimental resolution is plotted in the lower panel of figure~\ref{fig2}, as well as all the partial contributions.
The agreement is fairly good since each of the three main peaks is reproduced within an error of 2 meV, that can be partially attributed to the difference between the rare earth element of the respective study (Nd versus La).
Qualitatively, our data also seem to be in agreement with previous inelastic neutron scattering measurements~\cite{Qiu_PRB2008}.

\begin{table}
\caption{\label{tab:param} Weighting factors for the partial density-of-states}
\begin{ruledtabular}
\begin{tabular}{|c|c|c|c|c|}
 Atom & rel. $f_n(\langle Q \rangle)$ & $M_n$ & $2W_n = U_{iso}\langle Q \rangle^2$ & Weighting factor\\
\hline
La & 30.3 & 138.9 & 0,25798& 5.107\\
\hline
Fe & 11.46 & 55.8 & 0.2818 & 1.7756\\
\hline
As & 16.95 & 74.9 & 0,31752 & 2.792\\
\hline
O & 2.335 & 15.99 & 0,31752 & 0.248\\
\hline
\end{tabular}
\end{ruledtabular}
\end{table}

\begin{figure}[ptbh]
\begin{center}
\includegraphics[width=1.05\columnwidth]{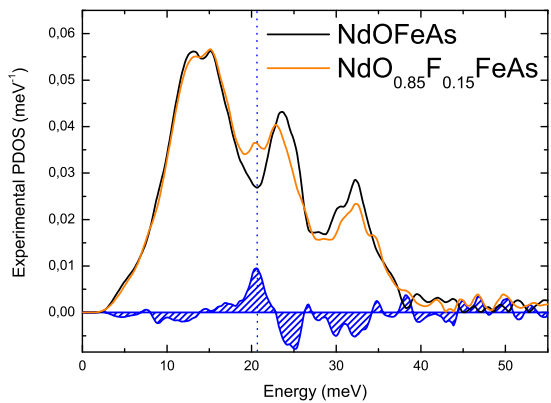}
\end{center}\vspace{-5mm}
\caption{(Color online) Evolution of the PDOS of NdOFeAs under F:O substitution. The area of both spectra have been normalized to unity. The blue line corresponds to the difference between the two spectra.}
\label{fig3}
\end{figure}

In Figure~\ref{fig3}, we show the evolution of the PDOS of NdOFeAs on substituting O by F. The overall shape of the spectra is conserved, except for the second peak. Indeed, we notice a striking change below 24 meV, with the appearance of an extra contribution to the PDOS around 21 meV. 
In order to rule out any effect of the energy resolution or the Q-sampling, we recorded the relevant part of the PDOS with a higher energy resolution of 1.8 meV (using the silicon (11-11-11) configuration at 21.747 keV (not shown here). The effect was still clearly visible. 

\par
Simulations of the substitution of O by F using the virtual crystal approximation (VCA) show that the PDOS obtained for the doped sample should \textit{a priori} not display any significant differences with respect to the one of the parent compound~\cite{Boeri}. From the calculation, three modes are composing the middle band extending from 20 to 28 meV. 
These three modes are Raman-active, as reported by Hadjiev and collaborators on undoped REOFeAs (RE = (La, Sm))~\cite{Hadjiev_PRB2008}. We emphasize that recent measurements in CeOFeAs~\cite{Chen_PRL2008} and in NdO$_{0.82}$F$_{0.18}$FeAs~\cite{Dubroka_PRL2008} clearly show the absence of any infrared-active features between 110 cm$^{-1}$ (13.6 meV) and 250 cm$^{-1}$ (31 meV). We thus naturally attribute the 21 meV bump in the doped spectra to a softening of at least 2 meV of one of the three Raman-active branches on doping.
Following Ref.~\onlinecite{Hadjiev_PRB2008}, these modes are mainly be attributed to respectively out-of phase c-axis polarized vibrations of the rare earth (with the A$_{1g}$ symmetry), As (A$_{1g}$ symmetry) and Fe (B$_{1g}$ symmetry). 

\begin{figure}[ptbh]
\begin{center}
\includegraphics[width=1.05\columnwidth]{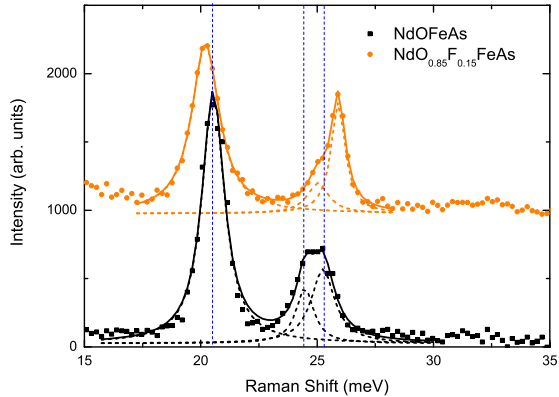}
\end{center}\vspace{-5mm}
\caption{(Color online) Raman scattering spectra of NdOFeAs (black points) and NdO$_{0.85}$F$_{0.15}$FeAs (red points). The dotted lines are the individual fits for each of the three phonon lines and the solid line is the sum of the fitted lines.}
\label{fig4}
\end{figure}

In order to gain further insight, and to see if one can explain the previous observation by the softening of the Raman-active zone center phonons, we have performed a (unpolarized) micro-Raman scattering experiment on various individual grains of the powders of doped and undoped samples. 
The experiment was carried out under a microscope (x100 magnification) attached to a LABRAM Jobin-Yvon Raman spectrometer (1800/mm gratings) operating in backscattering geometry, at 514.52 nm with an incident laser power density of about $5\times10^4~W/cm^2$.
The results shown in Figure~\ref{fig4} are in full agreement with those previously obtained by Hadjiev and collaborators~\cite{Hadjiev_PRB2008} on LaOFeAs and SmOFeAs.
We can clearly distinguish 2 peaks in the Raman response of the parent compound. The first one is located around 20.5 meV (166 cm$^{-1}$) and corresponds to the A$_{1g}$ mode of Nd.
This frequency is close to the one expected from the dependence of the phonon frequency on atomic masses and nearest neighbor distance given in Refs.~\onlinecite{Hadjiev_PRB2008, Atanassova_PRB94}. The distance Nd-O is found to be 2.3149 \AA~in pure NdOFeAs~\cite{Bos_condmat}. The expected renormalization of the rare earth A$_{1g}$ phonon with respect to LaOFeAs, given by $\omega_{Nd} / \omega_{La} \sim \sqrt(m_{La}d_{La-O}^3/(m_{Nd}d_{Nd-O}^3)) \sim$ 1.009, is very close to the experimentally observed one $\omega_{Nd} / \omega_{La} = $1.022. The same approach for Sm gives $\omega_{Nd} / \omega_{Sm} =$ 0.984 for an observed value of 0.976.
The second peak around 25 meV (200 cm$^{-1}$) is composed of two contributions\cite{Hadjiev_PRB2008} respectively arising from the As A$_{1g}$ mode (24.5 meV/197 cm$^{-1}$) and from the Fe B$_{1g}$ mode (25.3 meV/204 cm$^{-1}$).
As 15\% of O is substituted by F in NdOFeAs, an increase of $d_{Nd-O}$ of about 0.0113 \AA~is found~\cite{Bos_condmat}. Following the above arguments, this would lead to a very small softening of $\sim$ 0.15 meV of the Nd A$_{1g}$ phonon, in good agreement with the experimental finding of a softening of about 0.3 meV for this line.
Similarly, a small softening of the two other branches ($\sim$ 0.3 meV) would be expected from the small increase of $d_{Fe-As}$ in the doped compound but experimentally, a \textit{hardening} of about 0.6 meV of these two branches is found.

\par
According to recent dynamical mean field theory (DMFT) calculations on LaOFeAs, the parent compound behaves like a bad semiconductor, with a small gap in the electronic density-of-states, separating the quasiparticle peak from the Fermi level~\cite{Haule_PRL2008}. Upon F-doping, simulated in the VCA, the quasiparticle peak gets closer to the Fermi level, and the doped compound becomes more and more metallic. As a consequence, Coulomb screening is modified and one can expect variations of the frequency of phonon branches coupled to the electronic continuum. This effect will be likely most pronounced for the phonon involving the motion of atoms of the conducting FeAs layer since most of the states close to the Fermi level originate from the Fe 3$d$ bands. This could possibly explain the zone center hardening of the FeAs layer phonons found in our Raman measurement, unexpected from sole changes in lattice parameters.
On the other hand, the zone center softening of the Nd A$_{1g}$ mode seems consistent with these changes, but may be too small to be held responsible for the softening effect observed in the PDOS under F:O substitution. 
The latter is certainly related to electron-phonon coupling in oxypnictides and might then be attributed to a phonon softening at finite momentum transfer Q. Further experiments on single crystals, as well as lattice dynamics calculations taking into account strong electronic correlation effects, are thus required to understand this softening in the PDOS. Combined together, they will shed new light on the impact of electron-phonon coupling on the already rich physics of oxypnictides.

\section*{Note}
After first submitting this paper, we became aware of similar PDOS neutron~\cite{McQueeney}, x-ray~\cite{Baron}, as well as nuclear resonant~\cite{Higa} inelastic scattering measurements on iron based oxypnictide superconductors. 

\section*{ACKNOWLEDGEMENTS}
We are very grateful to the staff of the Swiss-Norwegian CRG Beamline (ESRF) for XRD measurements and J. Kreisel for providing access to his Raman setup. We acknowledge A. Walters for his critical reading of the manuscript.

\end{document}